\documentclass[aps,twocolumn,showpacs,superscriptaddress,10pt,prl]{revtex4-1}

\usepackage[T1]{fontenc}

\usepackage[colorlinks,bookmarks=false,citecolor=blue,linkcolor=red,urlcolor=blue]{hyperref}
\usepackage{amsmath}
\DeclareMathOperator{\tr}{Tr}
\usepackage{amsfonts}
\usepackage{amssymb}
\usepackage{graphicx}
\usepackage{color}
\usepackage{soul}
\usepackage{xcolor}
\usepackage{footmisc}
\usepackage{braket}

\allowdisplaybreaks

\begin{document}

\title{The effect of active photons on dynamical frustration in cavity QED}

\begin{abstract}

We study the far-from-equilibrium dynamical regimes of a many-body spin boson model with disordered couplings  relevant for cavity QED and trapped ions experiments, using the discrete truncated Wigner approximation (DTWA). 
We focus on the dynamics of spin observables upon varying the disorder strength and the frequency of the photons, finding that the latter can considerably alter the structure of the system's dynamical responses. When the photons evolve at a similar rate as the spins, they can induce  qualitatively distinct frustrated dynamics characterized by either logarithmic or algebraically slow relaxation. {The latter illustrates resilience of glassy-like dynamics in the presence of active photonic degrees of freedom, suggesting
 that disordered quantum many body systems} with resonant photons or phonons can display a rich diagram of non-equilibrium responses, with near future applications for   quantum information science. 

 \end{abstract}

\author{Shane P. Kelly}
\email[Corresponding author:~]{shakelly@uni-mainz.de}
\affiliation{Institut f\"ur Physik, Johannes Gutenberg Universit\"at Mainz, D-55099 Mainz, Germany}
\author{Ana Maria Rey}
\affiliation{JILA, NIST, Department of Physics, University of Colorado, Boulder, CO 80309, USA}
\affiliation{Center for Theory of Quantum Matter, University of Colorado, Boulder, CO 80309, USA}
\author{Jamir Marino}
\affiliation{Institut f\"ur Physik, Johannes Gutenberg Universit\"at Mainz, D-55099 Mainz, Germany}

\maketitle
\emph{Introduction ---} When coupled to a reservoir, a quantum system can undergo drastic modifications both of its static and dynamical features:
 both local~\cite{RevModPhys.59.1,vojta2006impurity} and global dissipation~\cite{mitra2006nonequilibrium, PhysRevB.76.195316, dalla2012dynamics, sieberer2013dynamical, gagel2014universal, tonielli2019orthogonality} can drastically morph the nature of quantum critical points, while non-ergodic systems can enter a regime of facilitated relaxation by coupling to a bath.
 Noticeable mechanisms include frustration relief by phonons in solids~\cite{ashcroft2010solid}, melting of MBL insulators coupled to a bath~\cite{nandkishore2014spectral,banerjee2016variable,kelly2020exploring}, damping of non-equilibrium superconductivity~\cite{mitrano2016possible, babadi2015far,kennes2017transient,sentef2016theory} or of pumped magnons~\cite{demokritov2006bose,PhysRevLett.108.246601,bender2014dynamic}. 

For a technological perspective, the dissipative and decohering effects of a bath are usually detrimental and pose an obstacle in developing  quantum mechanical devices~\cite{ladd2010quantum,bruzewicz2019trapped}.
On the other hand,   in the field of quantum simulation, a phononic or photonic bath can become a resource and a medium to engineer effective interactions~\cite{georgescu2014quantum}.
Examples include tunable long-rang interacting spin chains~\cite{britton2012engineered, zhang2017observation}, unidirectional photonics~\cite{lodahl2017chiral}, exotic tree-like interactions~\cite{bentsen2019treelike}, topological spin models~\cite{hung2016quantum,kim2020quantum} and frustrated magnets~\cite{tsomokos2008fully,grass2016quantum, gopalakrishnan2011frustration, marsh2020enhancing,strack2011dicke}, among the others. 
This success in interaction engineering is due to the ability to operate in a limit in which the bath responds faster than the system, and it becomes a  conduit for the transfer of many-body excitations.
Nevertheless, modern platforms ranging from trapped ions simulators~\cite{bruzewicz2019trapped, safavi2018verification, britton2012engineered, bohnet2016quantum,kim2010quantum, zhang2017observation} to cavity QED~\cite{baumann2010dicke,landig2016quantum, davis2019photon, kroeze2018spinor,norcia2018cavity,ritsch2013cold,PhysRevX.8.011002,kollar2017supermode,samutpraphoot2020strong,PhysRevLett.125.060402,rylands2020photon} and superconducting quantum circuits~\cite{blais2004cavity,houck2012chip} can operate in regimes where the excitations of the quantum environment (phonons or cavity photons) can resonate with the constituents of the system.
This opens the possibility for observing and simulating new physics when bath and system degrees of freedom can strongly couple and hybridize~\cite{PhysRevA.85.042114,PhysRevA.94.053637,damanet2019atom,halati2020numerically}.

\begin{figure}[b!]
     \includegraphics[width=0.48\textwidth]{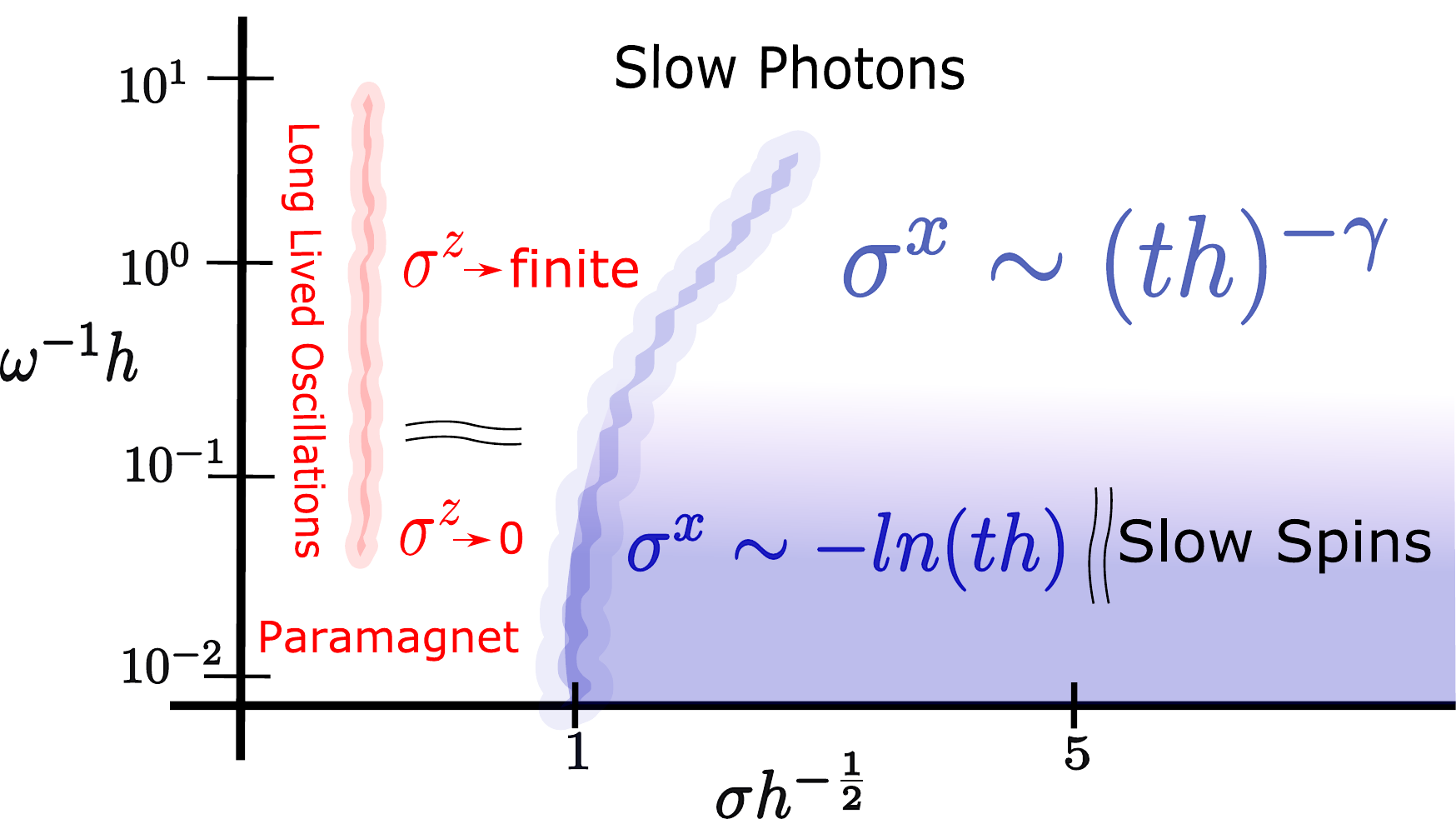}
     \caption{{Qualitative portrait}  of the dynamical responses at large $\alpha=N_b/N_s $ { and for an initial state with spins polarized in the $\hat{x}$ direction. The axes give }the photons frequency {$\omega h^{-1}$ and disorder strength $\sigma h^{-1/2}$, and with scales set by the transverse field $h$}.
         {At $\sigma h^{-1/2}>1$ and for large $\omega h^{-1}$}
         the spin magnetization experiences logarithmic  relaxation due to the fast photons mediating a long-range frustrated interaction.  {Upon reducing $\omega/ h$}, photons become active and the slow relaxation of the spins follows an algebraic decay. 
        { Large values of $\sigma$   'freeze' the dynamics of spins and their relaxation becomes critically slow for our simulations.}
     For weak disorder we observe dynamical paramagnetism and  a photon-assisted relaxation mechanism: the asymptotic transverse magnetization crosses over from a vanishing to a finite   value upon reducing  $\omega/ h$.  }
     \label{fig1}
 \end{figure}
 In this work we focus on the quenched dynamics of a many-body spin-boson quantum simulator with disordered couplings of variance $\sigma$. The model has recently attracted considerable interest as   a paradigmatic instance of spin glasses~\cite{gopalakrishnan2011frustration,strack2011dicke,buchhold2013,PhysRevA.86.023604} and of quantum associative memories~\cite{gopalakrishnan2011frustration,PhysRevB.91.014415, rotondo2015dicke, grass2016quantum, fiorelli2020signatures,fiorelli2020dynamics,marsh2020enhancing}. 
 In these works the bosons are treated in a regime where they respond much faster than the spins, and mediate long-range interactions frustrated by disorder:
   dynamics can show both exponential fast relaxation similar to a paramagnet~\cite{biroli2015,PhysRevLett.105.220401} and logarithmically slow dynamics akin to glassy systems~\cite{binder1986,biroli2015,cugliandolo1999,fisher1988b,marinari1998,montemurro2000,vincent1997}.
    {Here, we further investigate how  these dynamics are affected when the frequency of the photons (or phonons in a trapped ion implementation) is reduced and they can actively participate in spin relaxation. We find} that the now dynamically active photons retain frustrated features but of a qualitatively different nature signaled by algebraic relaxation in spin observables. This suggests that glassy-like dynamics can be robust to finite frequency bosons in disordered trapped ions or cavity QED many-body simulators.
     Furthermore, by controlling a frustration parameter (the ratio between the number of spins and photonic modes) we are able to relax both types of slow dynamics as illustrated in Fig.~\ref{fig3}.\\

     \emph{The model ---} We consider $N_b$ bosonic modes with annihilation operators $a_\lambda$ and $N_s$    {two level atoms with {with an energy splitting tunable by a transverse field} $h$; the latter are described by spin-half operators ${\sigma_i}^{x,y,z}$  with $i=1,2,...,N_s$}. We study their evolution under the hamiltonian~\cite{gopalakrishnan2011frustration,marsh2020enhancing,strack2011dicke} 
\begin{eqnarray}\label{ham}
    H&=&\sum_{j,\lambda} g_{j\lambda} \sigma_j^x(a_\lambda+a^\dag_\lambda)+h\sum_j\sigma^z_j+\sum_\lambda \omega_\lambda a^\dag_\lambda a_\lambda 
\end{eqnarray}
where we take $\omega_{\lambda}=\omega$ for all the bosons, and work in units of $\hbar=1$. 
We choose couplings $g_{j\lambda}$ from a random gaussian distribution with zero mean $\left<g_{jk}\right>=0$ and variance
\begin{eqnarray}
    \left<g_{kl}g_{k'l'}\right>-\left<g_{kl}\right>\left<g_{k'l'}\right>&= &\delta_{k,k'}\delta_{l,l'}\frac{\sigma^2\omega}{2N_t}
\end{eqnarray}
where the scaling by $N_t=N_s+N_b$ ensures the extensivity of the hamiltonian, $\sigma^2$ has units of frequency and the choice $\left<g_{jk}\right>=0$ ensures that the bosons do not super-radiate~\cite{strack2011dicke}.
This hamiltonian is relevant to   trapped ions   and cavity QED experiments~\cite{tsomokos2008fully}; here we will focus on the latter where the bosonic modes are cavity photons.
In these experiments, the couplings $g_{j\lambda}$ are proportional to the amplitude of the photon~(boson) mode $\lambda$ at the location of the $j^{th}$ atom~(spin), with the latter pinned at a given location in space~\cite{guo2019sign,PhysRevX.8.011002}.

In the limit of large photons' frequency    {$\omega\gg h$}, $\sigma^2$ the photons can be adiabatically eliminated~\cite{gopalakrishnan2011frustration,strack2011dicke}, and the hamiltonian~\eqref{ham} reduces to the following random transverse field Ising model
\begin{eqnarray}
    H_a=-\sum_{jk\lambda}\frac{g_{j\lambda}g_{k\lambda}}{\omega_\lambda}\sigma^x_j\sigma^x_k+h\sum_j\sigma^{z}_j.
    \label{eq:adiabadicham}
\end{eqnarray}
For $\omega_{\lambda}=\omega$, Eq.~\eqref{eq:adiabadicham} has been studied as a quantum version~\cite{nishimori1996,gopalakrishnan2011frustration} of the Hopfield model and shows both thermal and quantum spin glass phase transitions~\cite{hopfield1982, amit1985storing, amit1985a, hopfield1982, nishimori1996}.
For a sufficiently large ratio of bosonic modes to spins, $\alpha={N_b}/{N_s}$, the spin glass undergoes a quantum phase transition to a paramagnetic phase at a critical disorder strength $\sigma_c^2/h\approx 1$ \cite{nishimori1996, strack2011dicke}.
On the other hand, when the system is in the classical limit ($h/\sigma^2\ll1$), $\alpha$ acts as a frustration parameter, interpolating between a mostly separable interaction with ferromagnetic like ground states at small $\alpha$~\cite{amit1985storing}, to a model with a spin glass ground state at large $\alpha$~\cite{sherrington1975a,gopalakrishnan2011frustration,strack2011dicke}.
Both $\alpha$ and $h/\sigma^2$ are controllable in multi-mode cavity QED experiments~\cite{marsh2020enhancing}, and in the following we use both as tunable knobs to cross over   different dynamical regimes.

Distinctly from the previous body of literature~\cite{fiorelli2020signatures,fiorelli2020dynamics,grass2016quantum,rotondo2015dicke,carollo2020,fiorelli2019,rotondo2018,strack2011dicke,buchhold2013,PhysRevA.86.023604}, we consider real-time dynamics in regimes where it is not possible to separate the energy scales in Eq.~\eqref{ham}.
Spins are initialized in a completely polarized state along the $\hat{x}$-direction, and photons are initialized in their vacuum state.
Such a state can be achieved in cavity QED experiments by optically pumping the spins into an eigenstate of $\sigma^z$, followed by a $\pi/2$ pulse.
Our choice is motivated by interest in slow frustrated relaxation dynamics occurring after a quench in a strong field cooled spin glass~\cite{binder1986}.

 \begin{figure*}
     \includegraphics[width=0.3\textwidth]{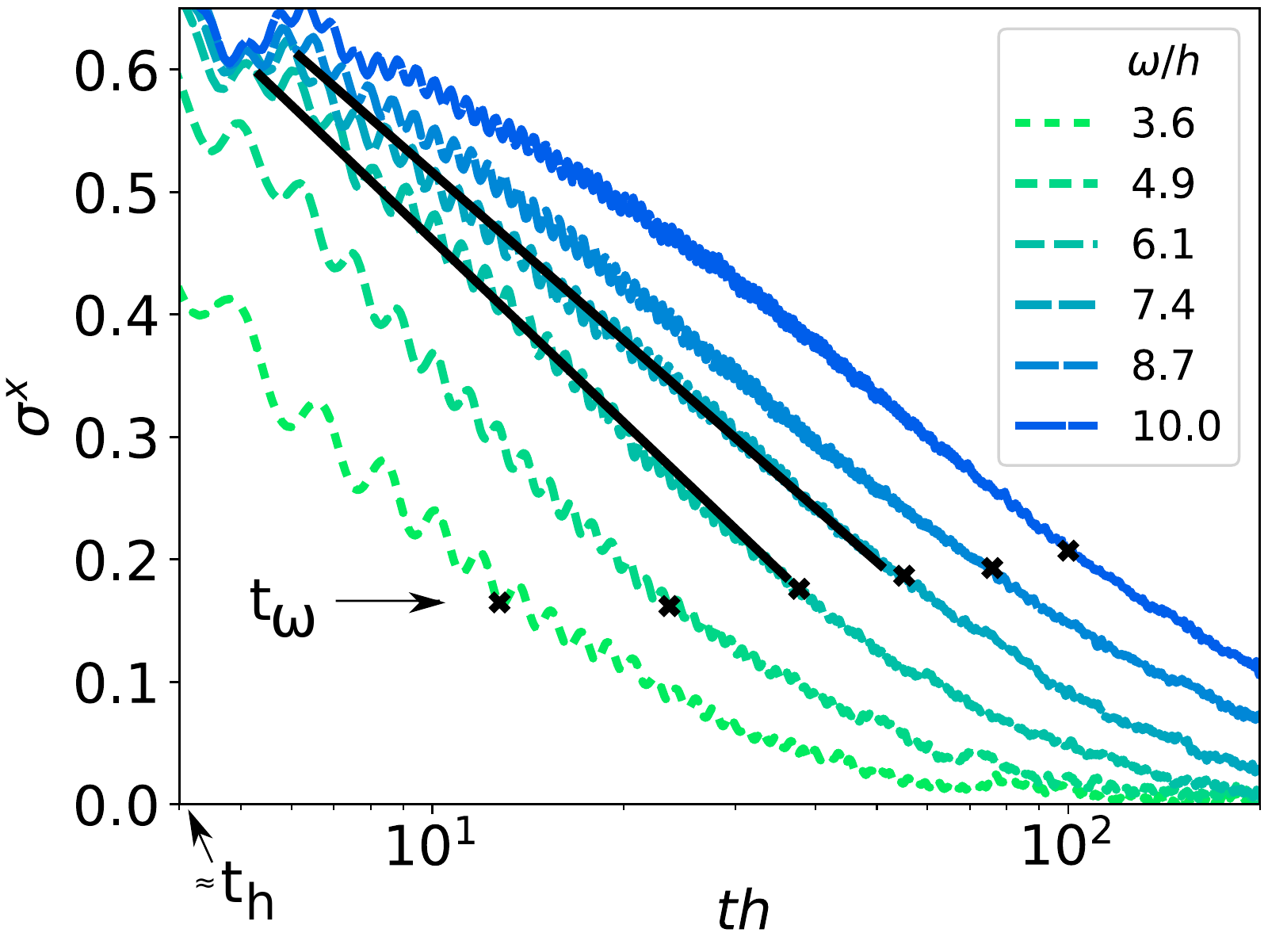}
          \includegraphics[width=0.305\textwidth]{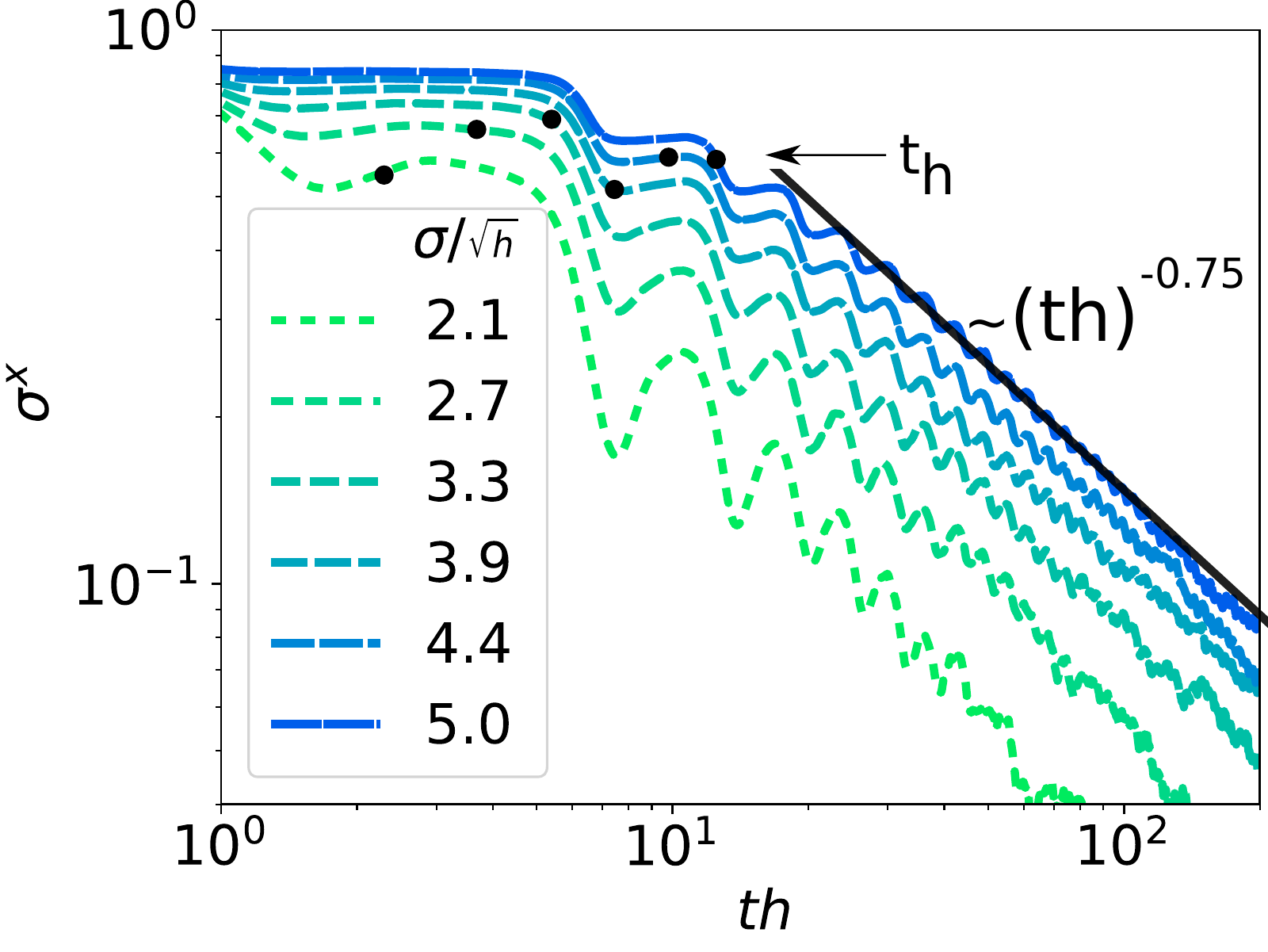}
               \includegraphics[width=0.3\textwidth]{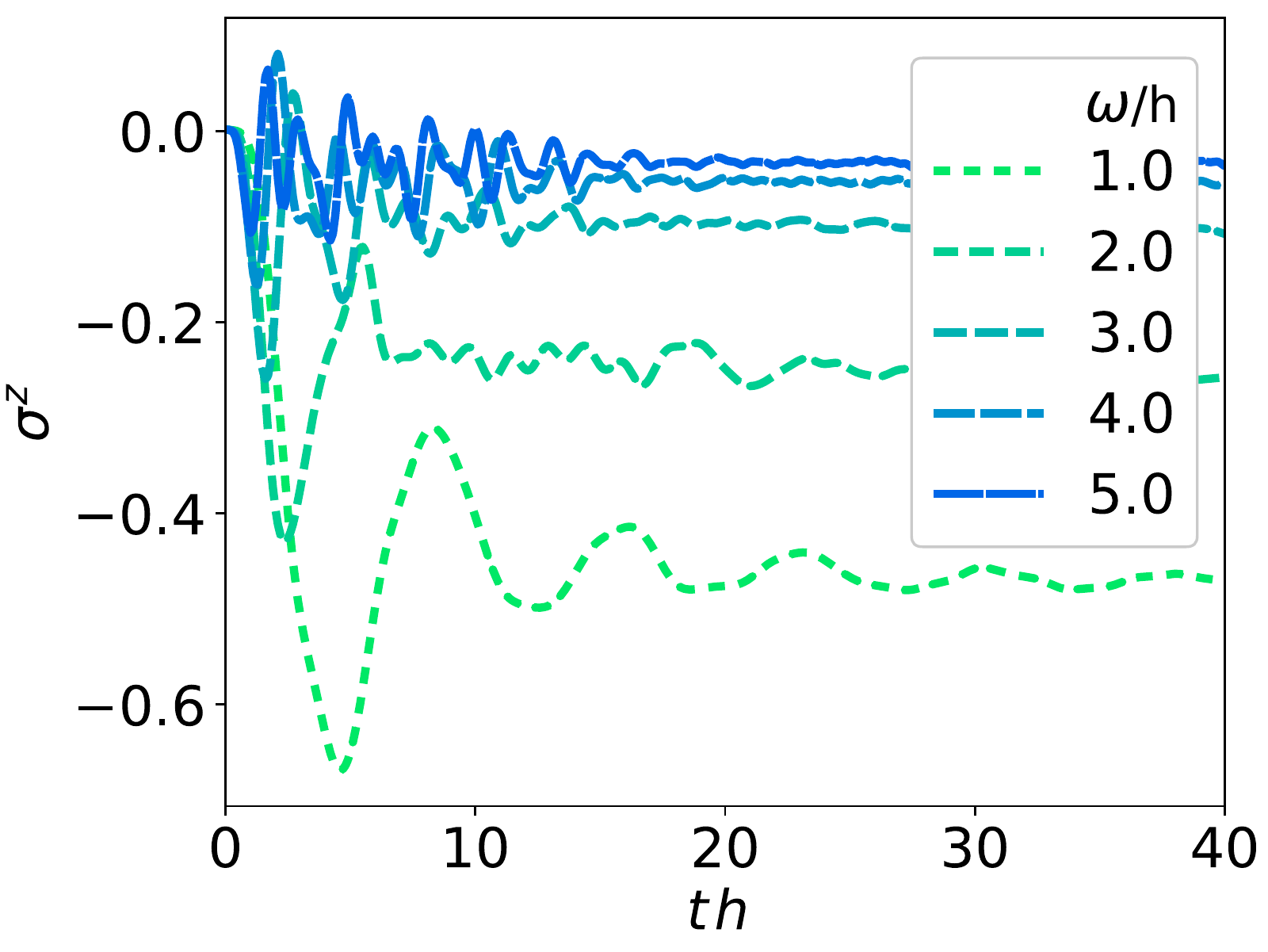}
     \caption{   {\textbf{Left panel}: \emph{Effect of photon frequency on frustrated dynamics}.
             System's parameters are $\alpha=10$, $N_t=1000$,   and $\sigma/\sqrt{h}=2$;  $\omega/h$ varies from $3.6$ (dashed, bright green) to $10$ (solid, dark blue),  deep in the regime of validity of adiabatic elimination. Logarithmic relaxation is marked by black  lines for intermediate values of   photon frequencies; for $\omega/h>7.5$ relaxation is slower. 
           Here dynamics are beyond the validity of the perturbative regime which holds up to $t\lesssim t_h\approx2/h$. At a second   timescale $t_\omega$ (marked by black crosses) logarithmic  relaxation crosses over into a qualitatively different dynamical regime.    \textbf{Central panel}: \emph{Dynamics beyond adiabatic elimination}.  The figure  shows  power law relaxation for $\omega/h=1$, $\alpha=10$, $N_t=1000$,  and $\sigma/\sqrt{h}$ varying from $2.1$~(dashed, bright green) to $5$~(solid, dark blue). The power law relaxation $t^{-\gamma}$ (with $\gamma\approx0.75$) starts after $t_h$ when perturbation theory breaks down (marked by the black dots); for the parameters considered in this panel, $t_\omega$ occurs before $t_h$. \textbf{Right panel}: \emph{Photon assisted relaxation in the dynamical paramagnet}. Dynamics of $\sigma^z$ after a quench into the paramagnetic phase for $\alpha=10$, $\sigma/\sqrt{h}=0.7$ and with $\omega/h$ shown in the legend. }  }
     \label{fig2a}
 \end{figure*}

 {\emph{Methods --- }} Since we study quench dynamics for any $\omega$ and $h$, we can not immediately work in the perturbative limit of Refs.~\cite{fiorelli2020signatures,fiorelli2020dynamics,marsh2020enhancing,grass2016quantum,rotondo2015dicke,carollo2020,fiorelli2019,rotondo2018}, or in the large $N$ semi-classical limit of Refs.~\cite{strack2011dicke,buchhold2013,PhysRevA.86.023604}.
 Instead, we employ a Discrete Truncated Wigner Approximation (DTWA)~\cite{schachenmayer2015a, orioli2017,davidson2017semiclassical,acevedo2017exploring,sundar2019analysis,zhu2019b,kunimi2020,nagao2020}.
 DTWA generalizes the Truncated Wigner Approximation~(TWA)~\cite{polkovnikov2010phase} to spin-half systems, by using the Wooters discrete Wigner function~\cite{WOOTTERS19871} to represent the quantum state.
For initial product states, the discrete Wigner function is positive and it can be therefore efficiently sampled.
In this work, we also perform the TWA sampling procedure of the photonic Wigner function as outlined in Ref.~\cite{orioli2017}.
Each sample is then evolved according to the classical equations of motion of the spin-half hamiltonian, and time dependent observables are captured by averaging the observables over different samples.
The accuracy of TWA and DTWA depends on the relevance  of quantum jumps that occur at the next level in the approximation~\cite{polkovnikov2010phase}.
For long range models, such as the  cavity QED problem studied here, DTWA is accurate at long times~\cite{schachenmayer2015a,kunimi2020}, and it can capture dynamical phase transitions~\cite{khasseh2020discrete,perlin2020spin}, thermalization~\cite{lepoutre2019a} and quantum glass dynamics~\cite{pappalardi2019quantum}

\emph{Time scales --- }    {The non-equilibrium dynamics of our model can be understood from the interplay of two time scales: the time scale $t_h$ at which spin flips in the $\sigma^x$ basis, generated by the transverse field $h$, contribute non-perturbatively to the quantum evolution, and the time scale $t_\omega$ that signals when photons start to participate in the spin dynamics. In the following we briefly sketch their estimate, referring to the Supplemental Material (SM) for further details.}
 
{ For the evaluation of $t_h$}, {we first observe that in the limit $\sigma^2/h\rightarrow \infty$, $\sigma^x_i$ becomes a conserved quantity and  spins remain frozen with expectation value  $\langle\sigma^x_i\rangle=1$, for $i=1,2,..., N_s$.  Applying, then,} perturbation theory in small $h$,  we find   the time  at which the    {transverse field starts to generate spin flips} 
\begin{eqnarray} \label{eq:fgr1}
    t_h\simeq\frac{\sigma^2}{2h^2}\frac{\sqrt{\alpha+\alpha^2}}{1+\alpha}\simeq_{(\alpha\gg1)} \frac{\sigma^2}{2h^2}.
\end{eqnarray}
{After $t_h$, multiple spin flips affect dynamics, and we capture these processes with DTWA.

   {The second time scale  $t_\omega$ is estimated by considering the first correction to the photon dynamics in adiabatic elimination}
\begin{eqnarray}\label{tomega}
    a_\lambda(t)=-\frac{1}{\omega}\sum_{j}g_{j\lambda}\sigma^{x}_j(t)-\frac{i}{\omega^2}\sum_jg_{j\lambda}\partial_t \sigma_j^x(t)+..
\end{eqnarray}
Substituting     Eq.~\eqref{tomega} into the coupling term of Eq.~\eqref{ham}, the leading order correction   to the adiabatic eliminated hamiltonian reads
\begin{eqnarray}
    \delta H_a=-\frac{1}{\omega^2}\sum_{\lambda}g_{k\lambda}g_{j\lambda}\sigma_k^x(t)\textsl{Re}\{(i\partial_t \sigma_j^x(t))\}.
\end{eqnarray}
    {Considering the latter as a perturbation, one can estimate via elementary arguments (see SM for details)  the time at which corrections to adiabatic elimination become significant:}
 \begin{eqnarray}
    t_\omega=t_s\frac{\omega^2t_s}{h^2t_h}.
\end{eqnarray}
   {In the equation above  $t_s$ is the    rate of change of $\langle\sigma^x(t)\rangle$, which can be bound 
from below by $1/h$ and from above by $t_h$.}
For the lower bound, we estimate that $\langle\sigma^x(t)\rangle$ evolves fastest when the effective field produced by the photons is zero, and the spin precesses around $\sigma^z$.
While for the upper bound, we estimate that the slowest $\langle\sigma^x(t)\rangle$ can evolve (while remaining partially polarized) is determined by the perturbation theory estimate in Eq.~\eqref{eq:fgr1}.
From such bounds, we   infer ${\omega^2}/{(\sigma^2h^2)}\lesssim t_\omega\lesssim\omega^2\sigma^2/h^4$.
When $\sigma^2/h \sim 1$, the bounds on $t_\omega$ approach each other and we find $t_{\omega}\sim\omega^2/h^3$.

When {$\omega^2/h\gtrsim\sigma^2$}  we have the timescales' separation $t_\omega\gtrsim t_h$: in general, $t_\omega$ increases with the photons' frequency since it signals dynamical breakdown of the adiabatic elimination regime. Upon tuning $\omega$ we can tune the ratio between $t_h$ and $t_\omega$ and in the following we extract results in both regimes.\\

\begin{figure*}[t!]
     \includegraphics[width=0.32\textwidth]{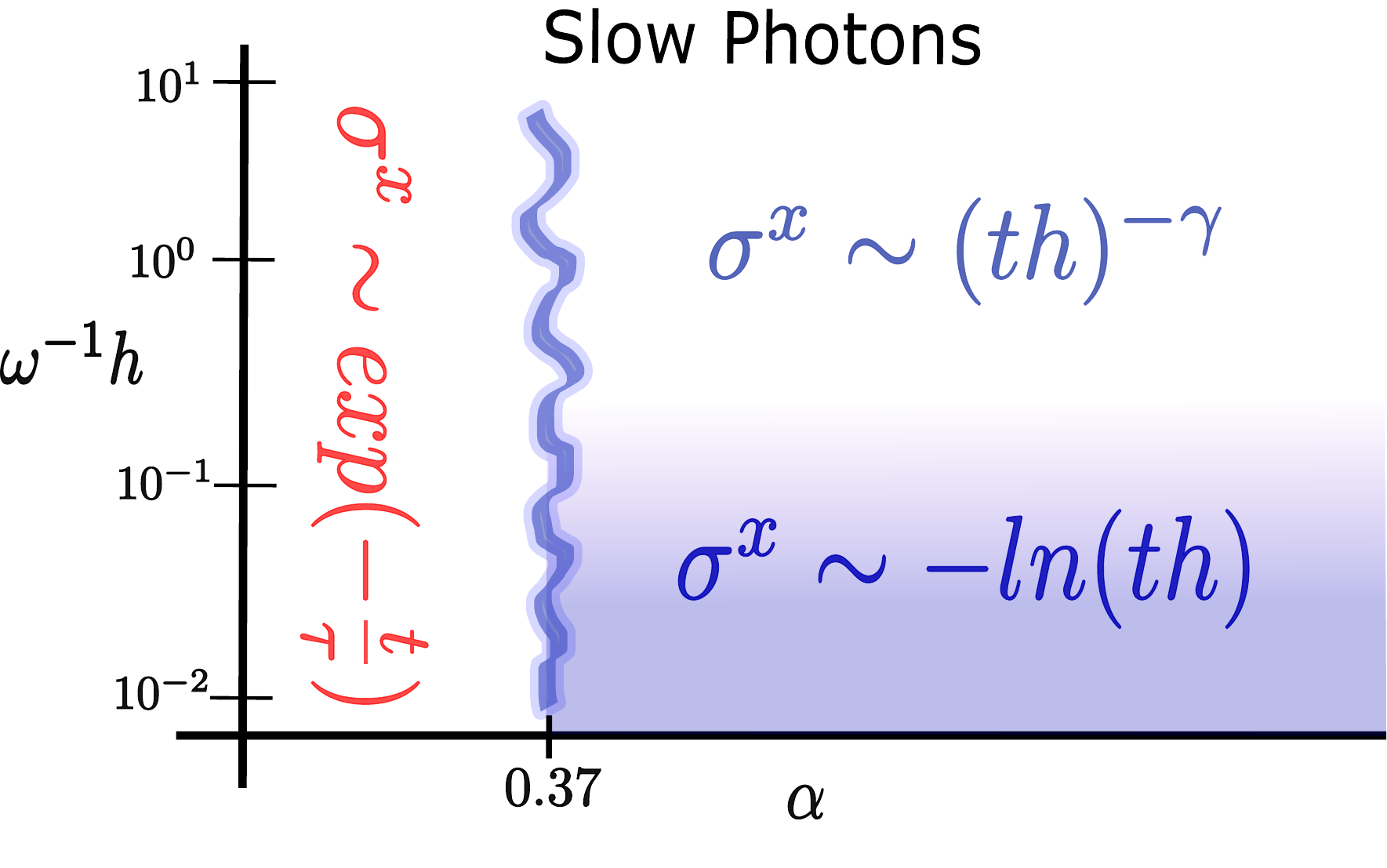}
     \includegraphics[width=0.3\textwidth]{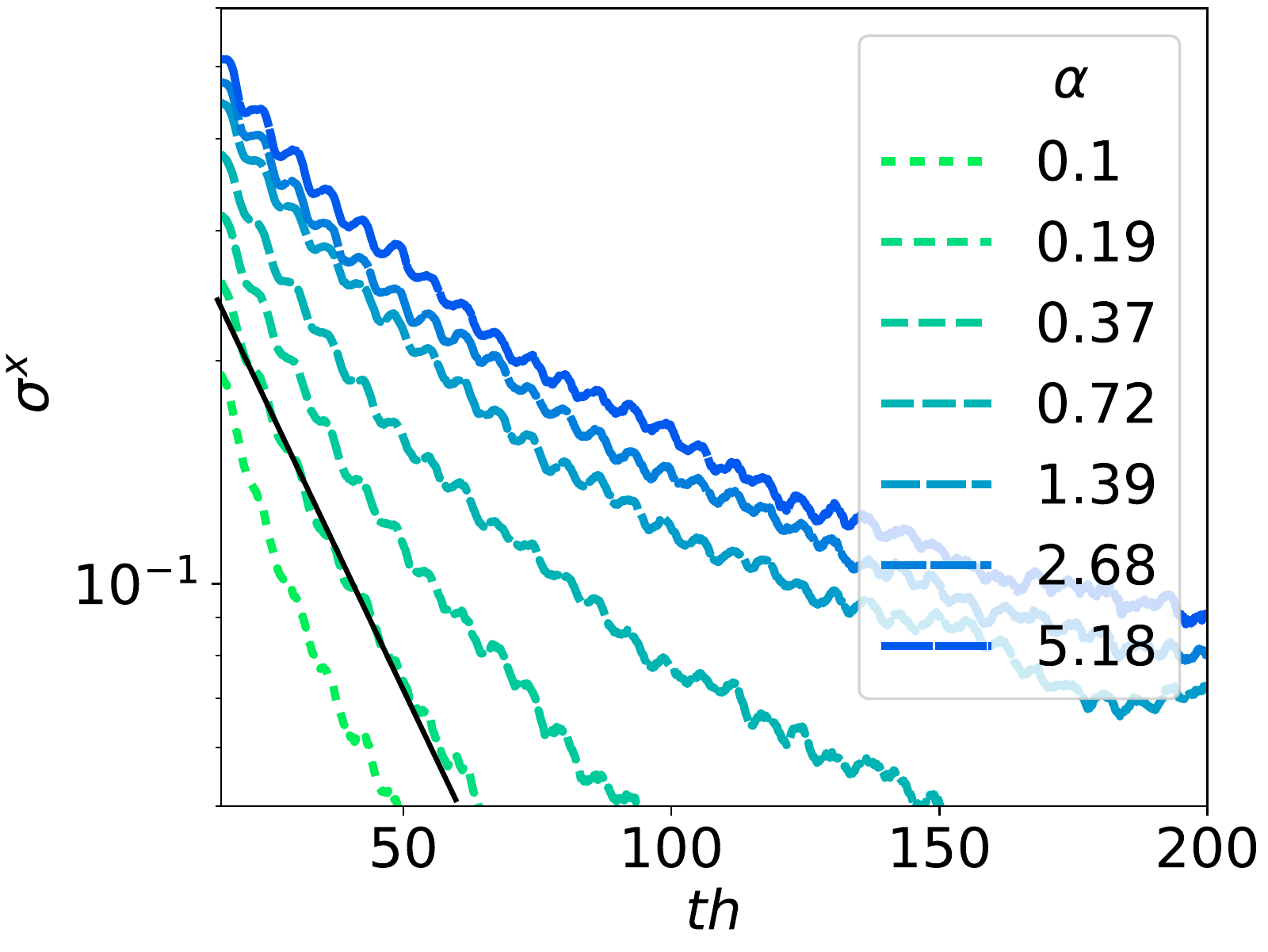}
     \includegraphics[width=0.3\textwidth]{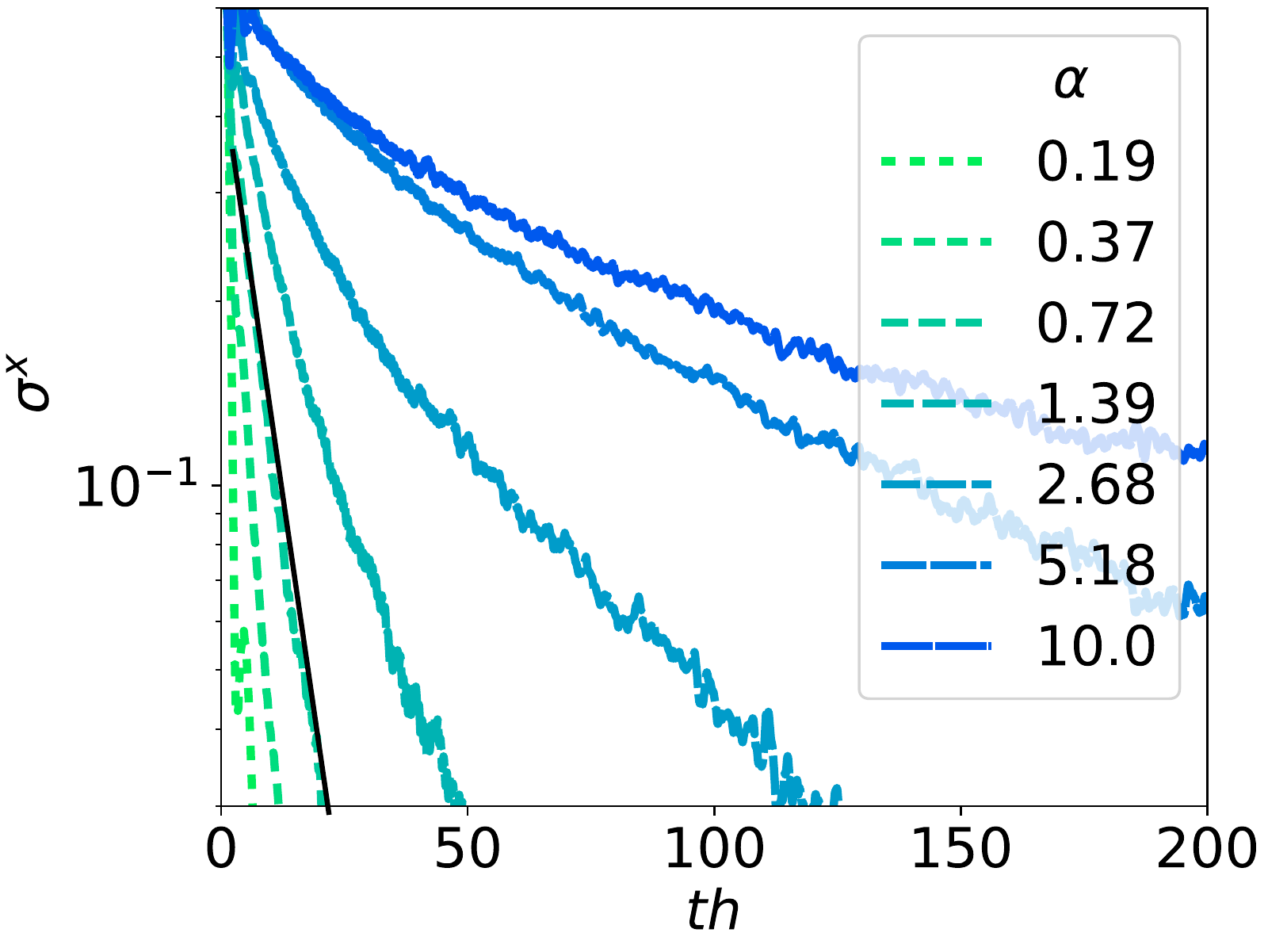}
     \caption{   {\textbf{Left panel}: diagram of dynamical responses as a function of $\alpha$  and $\omega^{-1}$ for $\sigma>\sigma_c$. The critical value of $\alpha$ in general depends on $\sigma$, the value $\alpha_c\approx0.37$ shown is for $\sigma/\sqrt{h}=5$. \textbf{Central panel}: transition to exponential relaxation at $\alpha\lesssim0.37$ (for $\sigma/\sqrt{h}=5$ and $\omega/h=1$). \textbf{Right panel}: transition from  sub-exponential to exponential relaxation for $\alpha\lesssim\alpha_c\approx3$  (here $\sigma/\sqrt{h}=2$ and $\omega/h=10$). System's parameters in all plots are  $N_t=640$; straight black lines mark the regimes of exponential relaxation in the central and right panels (the sampling error in $\sigma^x$ is around $\approx10^{-2}$, and we therefore do not display dynamics when $\sigma^x$ is below such threshold). In the SM we plot both figures in log-log scale.  }}
     \label{fig3}
 \end{figure*}
 
 \emph{Dynamical responses --- }
    {The quench dynamics of~\eqref{ham} organize into a rich set of responses comprising logarithmic, algebraic and exponential relaxation of spin observables, upon varying the disorder strength and the frequency of the photon, as summarized  in Fig.~\ref{fig1}.
 The model~\eqref{eq:adiabadicham} 
undergoes a quantum phase transition from a paramagnet to a glass~\cite{gopalakrishnan2011frustration,strack2011dicke}. Therefore,   we start   looking for signatures of dynamical paramagnetism~\cite{biroli2015,lerose2019impact} and glassy dynamics which is typically characterized by slow sub-exponential relaxation~\cite{binder1986,biroli2015,cugliandolo1999,fisher1988b,marinari1998,montemurro2000,vincent1997}.}

In the first  panel of Fig.~\ref{fig2a} we show that in the adiabatic limit ($\omega\gg h$) the relaxation of $\sigma^x(t)$ has logarithmic character.
In the same figure we illustrate that such logarithmic dynamics yields into another form of slow sub-exponential relaxation after $t_\omega$.
When $t\gtrsim t_\omega$ the role of photons cannot  be neglected and the effective model in~\eqref{eq:adiabadicham} ceases to describe the dynamics of the many-body spin-boson hamiltonian~\eqref{ham}.
 
 In the second panel of Fig.~\ref{fig2a} we present dynamics in the non-adiabatic limit where $\omega \approx h$. 
  {Despite the inability of the photons to mediate a static disordered interaction for the atoms as in the large $\omega$ regime (cf. with the  hamiltonian~\eqref{eq:adiabadicham}), a form of sub-exponential relaxation, reminiscent of the glass phase, persists after $t_h$ where perturbative dynamics do not hold anymore.    {In this parameter regime we have $t_\omega\lesssim t_h$ and therefore we do not observe any dynamical crossover.}}
The now active photons are able to partially relieve frustration of the  model and accelerate relaxation from a $\ln(ht)$ to a power law behavior $\sim t^{-\gamma}$ with $\gamma\simeq-0.75$ (see central panel of Fig.~\ref{fig2a}).
Since the magnetization does not relax in a exponential fashion as it occurs in the paramagnetic phase, this algebraic relaxation reveals a dynamical frustration present when photons actively participate in dynamics and constitutes one of the central results of our work. 
{In both  regimes of strong spin-boson coupling addressed in the left and central panels of Fig.~\ref{fig2a},  the decaying magnetization and dynamics of photons lock at a common frequency   $\propto \omega$.}


We now discuss the regime of dynamical paramagnetism.
For very small disorder ($\sigma^2/h\ll1$), the transverse field dominates spin dynamics, and spins simply precess around the $\hat{z}$-direction until disorder-induced dephasing relaxes them.
At moderate disorder strength, but still below the spin glass transition $\sigma_c^2\approx h$,  {inhomogeneous dephasing plays a significant role, and  magnetization swiftly finds a steady state ({on timescales $\propto 1/\omega$)}, as shown in right panel of Fig.~\ref{fig2a}.}
{In the adiabatic limit $\omega\gg h$, such steady state is completely depolarized} similar to other transverse field Ising models, where quenches from states completely polarized in the  $\sigma^x$ direction result in a compete loss of magnetization~\cite{PhysRevB.74.144423,biroli2015,lerose2019impact}. 
At smaller $\omega$ the photons exchange energy effectively with the spins and
this results in the latter relaxing towards a steady-state with finite $\sigma^z$ similar to a low temperature quench into a paramagnetic phase~\cite{PhysRevB.74.144423,biroli2015,lerose2019impact, kelly2019thermalization}.
The relaxation to a finite $\sigma^z$, as $\omega/h$ decreasing, occurs as a crossover and it is shown in the right panel of  Fig.~\ref{fig2a}.
{This can be captured by assuming the steady state is described by a thermal state with $H= h\sum_j\sigma^z_j+ \omega\sum_\lambda a^{\dagger}_{\lambda}a_{\lambda}$, and with an effective temperature set by the requirement that the energy of the thermal ensemble should match the energy of the initial state. This results in an effective temperature that decreases linearly with $\omega/h$ and a thermal spin polarization $\sigma^z$ consistent with the simulation results shown in the right panel of Fig~\ref{fig2a} (See SM for details)}

{\emph{Tuning the ratio of  spins and bosonic modes --- } So far we have discussed the dynamics at large values of $\alpha$. }
In the ground state of the classical Hopfield model (Eq.~\eqref{eq:adiabadicham} with $h=0$) reducing $\alpha$ relieves frustration~\cite{hopfield1982}.
Similarly, we observe here a crossover from sub-exponential  relaxation to an exponential relaxation by reducing $\alpha$ below some $\alpha_c$.
This is shown in Fig.~\ref{fig3}, where we have included a quench at large disorder and small photon frequency, and we show a crossover to exponential relaxation around $\alpha=0.37$ (central panel).

We can obtain an approximate estimate for the crossover $\alpha_c$    {as follows}.
If the system shows exponential relaxation,  we can assume glassy dynamic are not preventing thermalization, and the steady state will be described by a Gibbs ensemble~\cite{binder1986, deutsch2018}.
In the limit of small $h/\sigma^2$, and assuming that the photons are not condensed, the statistical Gibbs weight is associated to a classical Hopfield model coupled to $N_b$ non-interacting photons (see SM for details).
The classical Hopfield model has a   phase transition between a paramagnetic phase above a temperature $T_c = {\sigma^2N_s}(1 + \sqrt{\alpha})/{2N_t}$ and a variety of spin-glass states below $T_c$ with the spins freezing in random directions~\cite{amit1985storing}.
Thus, we expect that if the effective temperature of the final state is above $T_c$, we will see paramagnetic   relaxation, while below $T_c$, we might still observe glass like relaxation.
To compute the effective temperature of the initial state, we compare the   energy of the thermal states, $U(T)$ and the energy of the initial state $E$, and solve for the temperature $T=U^{-1}(E)$.
For the initial state discussed above (fully polarized along the $\hat{x}$-direction), we find a temperature  $T= {\sigma^2N_s}(1+\sqrt{3})/{4N_t}$, and therefore we expect a crossover from paramagnet like relaxation to sub-exponential relaxation around $\alpha_c\approx  \left(\sqrt{3}-1\right)/2\simeq0.366$.

This is confirmed in the central panel of Fig.~\ref{fig3}, where we find that above $\alpha\gtrsim0.37$, relaxation dynamics turn from exponential to sub-exponential.
For finite values of $h/\sigma^2$, quantum fluctuations correct the estimate given above and reduce the critical temperature~\cite{nishimori1996,mukherjee2015,goldschmidt1990}, with a resulting  increased  $\alpha_c$. An instance of this effect is illustrated in the right panel of Fig.~\ref{fig3} (see SM for the characterization of the sub-exponential relaxation in both regimes).

   {Following analogous arguments,  we  anticipate that sub-exponential relaxation will become exponential by increasing the energy of the initial state.
To confirm such an expectation, we perform a composite quench starting from a  fully polarized initial state along the $\hat{x}$-direction, and displacing at the same time  all the photons   by a uniform amplitude shift $\left<a(t=0)\right>= {2N_s}/{\sqrt{ N_t    {\omega/h}} }$.
In the SM we show that, in this case, all relaxation dynamics turn into exponential $\propto\exp(-t/\tau)$ when the energy of the initial state is sufficiently high.  The characteristic time $\tau$  increases with   $\sigma$ and decreases with  $\omega$.}\\

\emph{Photon losses --- } We now briefly comment on the effect of photon losses which is relevant for cavity QED experiments.
We modify our simulations by adding Langevin damping and noise terms to the photon dynamics~\cite{gardiner2004quantum,gelhausen2018,sels2020a,torre2013}, and focus our attention to the effect of photon loss on the sub-exponential relaxation.
We find (see SM) that the $\ln(ht)$ relaxation in the adiabatic limit remains for moderate loss, while for larger values of loss, relaxation accelerates.
This occurs because, in the adiabatic limit, the primary effect of dissipation is to reduce the effective field produced by the photons  and effectively move the adiabatic eliminated model  towards the paramagnetic regime.
For smaller $\omega$, the effect of photon loss also accelerates relaxation  as illustrated in the SM. \\
 
\emph{Perspectives --- }Our work illustrates that dynamics can display qualitative different features when neither photons~\cite{gopalakrishnan2011frustration,fiorelli2020signatures,fiorelli2020dynamics,marsh2020enhancing,grass2016quantum,sartori2015spin,rotondo2015dicke,carollo2020,fiorelli2019,rotondo2018,strack2011dicke,buchhold2013,PhysRevA.86.023604} nor atoms (as in photonic glasses~\cite{pierangeli2019large,ghofraniha2015experimental,tommasi2016robustness,moura2017replica,basak2016large, nixon2013observing}) can be adiabatically eliminated.
Future work might consider different inhomogeneous spin-boson couplings enabling investigations of other phenomena such as supersolidity and topological defects~\cite{gopalakrishnan2009emergent,gopalakrishnan2010atom}.
In such a framework, one could also access the fate of associative memory phases~\cite{gopalakrishnan2011frustration,strack2011dicke,PhysRevB.91.014415, rotondo2015dicke, grass2016quantum, fiorelli2020signatures,fiorelli2020dynamics,marsh2020enhancing} when photons cannot be integrated out. It is currently unclear whether photonic degrees of freedom could enhance memory retrieval or represent a hindrance. 
Other potential extensions include   the interplay of active photons with   multi-level   atoms~\cite{shchadilova2020fermionic}.

 Furthermore, the effects of active photons on scrambling of atoms in cavity QED simulators~\cite{PhysRevA.99.051803,PhysRevX.9.041011} (or trapped ions~\cite{garttner2017measuring,lewis2019unifying,PhysRevX.7.031011}) remain completely unexplored; this would represent a key future extension relevant for probing the dynamics of quantum information in open quantum systems.\\

 \emph{Acknowledgments-- }   S.~P.~K. and J.~M.~acknowledge stimulating discussions with M. Buchhold, G. Biroli, E. Demler, D. Sels, Y. Wang, G. Zarand.  We are indebted to M. Marcuzzi for his critical reading of our work. We acknowledge I. Kimchi, B. Lev and D. Young for their careful reading of the manuscript.  S. P. K. and J. M. acknowledge support by the Dynamics and Topology Centre funded by the State of Rhineland Palatinate and from the DFG through the SFB 'QuCoLiMa' (TRR306). A.~M.~R. acknowledges W911NF-19-1-0210, NSF PHY1820885, NSF JILA-PFCPHY-1734006 grants, QLCI-2016244 and by NIST. 

 \bibliography{cavity}
 \onecolumngrid
 
\appendix
\section{Thermodynamics for in the weak coupling limit}

{
At zeroth order in the weak coupling limit the thermal state is: 
\begin{eqnarray}
    \rho(\beta)=e^{-\beta (h\sum_i^{N_s} \sigma^z_i+\omega \sum_i^{N_b} n_i)}/Z
\end{eqnarray}
where $Z=\tr[\rho(\beta)]$.  The energy per spin of such a state is:
\begin{eqnarray}
    E(\beta)=-h \tanh(\frac{\beta h}{2})+\omega\alpha n(\beta)
\end{eqnarray}
where $n(\beta)=(e^{\beta\omega}-1)^{-1}$ is the Bose Einstein distribution function.
Since our initial state has energy $E=0$, the constraint $E(\beta)=0$ determines the temperature of the steady state and in turn the steady state polarization:
\begin{eqnarray}
    \lim_{t\rightarrow \infty}\sigma^z(t)\approx-\tanh(\frac{\beta h}{2})
\end{eqnarray}
Since $\tanh(\beta h/2) \in (0,1)$ for positive $\beta$, the constraint $E(\beta)=0$ sets an upper bound on bosonic occupation: $n(\beta)<h/(\omega \alpha)$. Using the Bose Einstein distribution function, this yields a constraint on the temperature:
\begin{eqnarray}
    \beta^{-1}<\frac{\omega}{\ln(\omega\alpha/h+1)}
\end{eqnarray}
and thus, for large alpha ($\alpha=10$ in our simulations), decreasing $\omega$ decreases the temperature and increases the $-\sigma^z$ polarization.  A numerical solution  of $E(\beta,\omega)=0$ for the effective temperature $\beta^{-1}$ finds the effective temperature decreases linearly with $\omega$ in the window of frequencies considered in the main text.  The spin polarization as a function of effective temperature is $\sigma^z=-\tanh(\frac{\beta h}{2})$ and is consistent with the steady state observed in simulations.
}

\section{Thermodynamics for the Classical Hopfield Model}
The partition function for Hamiltonian (1) in the main text becomes classical for $h=0$ and for high temperature photons.
In this limit, we study the classical partition function with Hamiltonian:
\begin{eqnarray}
    H=\sqrt{2}\sum_{j,\lambda} g_{j\lambda} \sigma_j^xx+\sum_\lambda \frac{\omega_\lambda}{2} (x_\lambda^2+p_\lambda^2) ,
\end{eqnarray}
$\sigma^x=\pm1$, and real classical coordinates $x=\sqrt{2}\Re a$ and $p=\sqrt{2}\Im{a}$.
For the initial state considered in the main text, $\sigma^x=1$ and $x=p=0$, yielding $\left<H\right>=0$.
To identify an effective temperature ($T=1/\beta$) for the steady state, we shift the boson fields and consider the model:
\begin{equation}\begin{split}
    H&=H_s+H_b=\\
    &=-\sum_{jk\lambda}\frac{g_{j\lambda}g_{k\lambda}}{\omega}\sigma^x_j\sigma^x_k +\frac{\omega}{2}\sum_\lambda(x^2_{\lambda}+p^2_{\lambda}),
\end{split}
\end{equation}
where $x_\lambda$ and $p_\lambda$ are the canonically conjugated coordinates of the bath. 
If we implement the disorder in $g_{j\lambda}$ with binary random variables, the spin part of the hamiltonian becomes the Hopfield model in different units $H_s=s H_h$, where $s= {\sigma^2 N_s}/{2N_t}$ and $H_h$ is the Hopfield hamiltonian~\cite{amit1985storing}.
In the paramagnetic phase, the Hopfield model has internal energy~\cite{amit1985storing}
\begin{eqnarray}
    U(\beta)=\frac{\alpha}{2} \left(1 -\frac{1 }{1-\beta }\right),
\end{eqnarray}
therefore, our model has internal energy per unit spin
\begin{eqnarray}
    U(\beta)=s\frac{\alpha}{2} \left(1 -\frac{1}{1-s\beta }\right)+\alpha \beta^{-1}.
\end{eqnarray}
Imposing $ U(\beta)=\left<H\right>=0$, we find   the effective temperature   in the main text.

\section{Time Scales}
To obtain $t_h$ in the main text, we apply perturbation theory on the bare hamiltonian $H$, and treat $V_h=h\sum_j \sigma^z$ as the perturbation to the hamiltonian
\begin{eqnarray}
    H_0&=&\sum_{j,\lambda} g_{j\lambda} \sigma_j^x(a_\lambda+a^\dag_\lambda)+\sum_\lambda \omega_\lambda a^\dag_\lambda a_\lambda ,
\end{eqnarray}
which has eigenstates labeled by the eigenvalues of $\sigma^x_j$: $\sigma^x_j\left|\{s_k\}\right>=s_j\left|\{s_k\}\right>$.
For each subspace given by the quantum numbers $\{s_j\}$ the photons can be diagonalized by a displacement operator.
Therefore the energies for the states labeled by $\{s_j\}$ and photon numbers $n_\lambda$ are:
\begin{eqnarray}
    H_0=-\sum_{jk\lambda}\frac{g_{j\lambda}g_{k\lambda}}{\omega_\lambda}s^x_js^x_k+\sum_\lambda \omega_\lambda n_\lambda.
\end{eqnarray}
Since $V$ can only flip one spin we will estimate $t_h$ from the time the perturbation $V$ will take to flip one spin.
Focusing on the $j^{th}$ spin, the energy difference after the flip is
\begin{eqnarray}
    \Delta E_j=-\sigma^2\sum_{k\lambda}\frac{\xi_{j\lambda}\xi_{k\lambda}}{2N_t}\sigma^x_k=-\sigma^2\sum_{k\lambda}\frac{\xi_{j\lambda}\xi_{k\lambda}}{2N_t},
\end{eqnarray}
where $\xi_{j\lambda}\sqrt{\sigma^2\omega/2N_t}=g_{j\lambda}$ such that $\xi_{j\lambda}$ has variance $1$.
The average magnitude of $\Delta E_j$ is given as:
\begin{equation}\begin{split}
    \left<(\Delta E_j)^2\right>&=\frac{\sigma^4}{4N_t^2}\sum_{kk'\lambda\lambda'}\left<\xi_{j\lambda}\xi_{j\lambda'}\xi_{k\lambda}\xi_{k'\lambda'}\right>=\\
    &=\frac{\sigma^4}{4N_t^2}(N_bN_s+N_b+N_b^2)\approx\sigma^4\frac{\alpha+\alpha^2}{4(1+\alpha)^2}.
    \end{split}
\end{equation}
This gives the energy denominator in perturbation theory, and we estimate:
\begin{eqnarray}
    t_h\simeq\frac{\sigma^2}{2h^2}\frac{\sqrt{\alpha+\alpha^2}}{1+\alpha}\simeq_{(\alpha\gg1)} \frac{\sigma^2}{2h^2}.
\end{eqnarray}

In the main text, we quoted the perturbation theory estimate $t_\omega$.
This is obtained by considering 
\begin{eqnarray}
    V_\omega =-\frac{1}{\omega^2}\sum_{\lambda}g_{k\lambda}g_{j\lambda}\sigma_k^x(t)\textsl{Re}\{(i\partial_t \sigma_j^x(t))\}
\end{eqnarray}
as a perturbation to the adiabatic eliminated hamiltonian $H_a$.
The Heisenberg equation of motion gives $i\partial_t\sigma_j^x\sim\sigma^y$, and for any eigenstate of the $\sigma^x_i$ operators, this perturbation can only flip one spin at a time.

Since the eigenstates of $H_a$ are analytically inaccessible, we focus on the limit when $h/\sigma^2$ is small and use 
\begin{eqnarray}
    H_0'=-\sum_{jk\lambda}\frac{g_{j\lambda}g_{k\lambda}}{\omega_\lambda}\sigma^x_j\sigma^x_k
\end{eqnarray}
to compute the energy denominator. This limit is appropriate for the spin glass region of parameter space when $\sigma^2>h$.
Therefore, we can again use the energy denominator $\Delta E_j$ calculated above.
To obtain the average magnitude of the perturbation $V_\omega$, we consider the action of $V_\omega$ on the fully polarized state, and estimate $\left|i\partial_t\sigma^x_j\right|\approx  {1}/{t_s}$.
This gives an estimate for the magnitude of the perturbation as:
\begin{eqnarray}
    V_\omega\approx\sum_{\lambda k}\frac{\sigma^2}{\omega}\frac{1}{t_s}\xi_{j\lambda}\xi_{k\lambda}.
\end{eqnarray}
The average magnitude squared is then:
\begin{eqnarray}
    \left<V^2_{\omega}\right>=4\frac{\sigma^4}{\omega^2t_s^2}\frac{\alpha+\alpha^2}{(1+\alpha)^2},
\end{eqnarray}
and perturbation theory then yields:
\begin{eqnarray}
    t_\omega\approx \frac{\sqrt{\left<(\Delta E)^2\right>}}{\left<V_\omega^2\right>}=t_s\frac{\omega^2t_s}{h^2t_h},
\end{eqnarray}
as in the main text.\\

 \begin{figure}[h!]
     \includegraphics[width=0.35\textwidth]{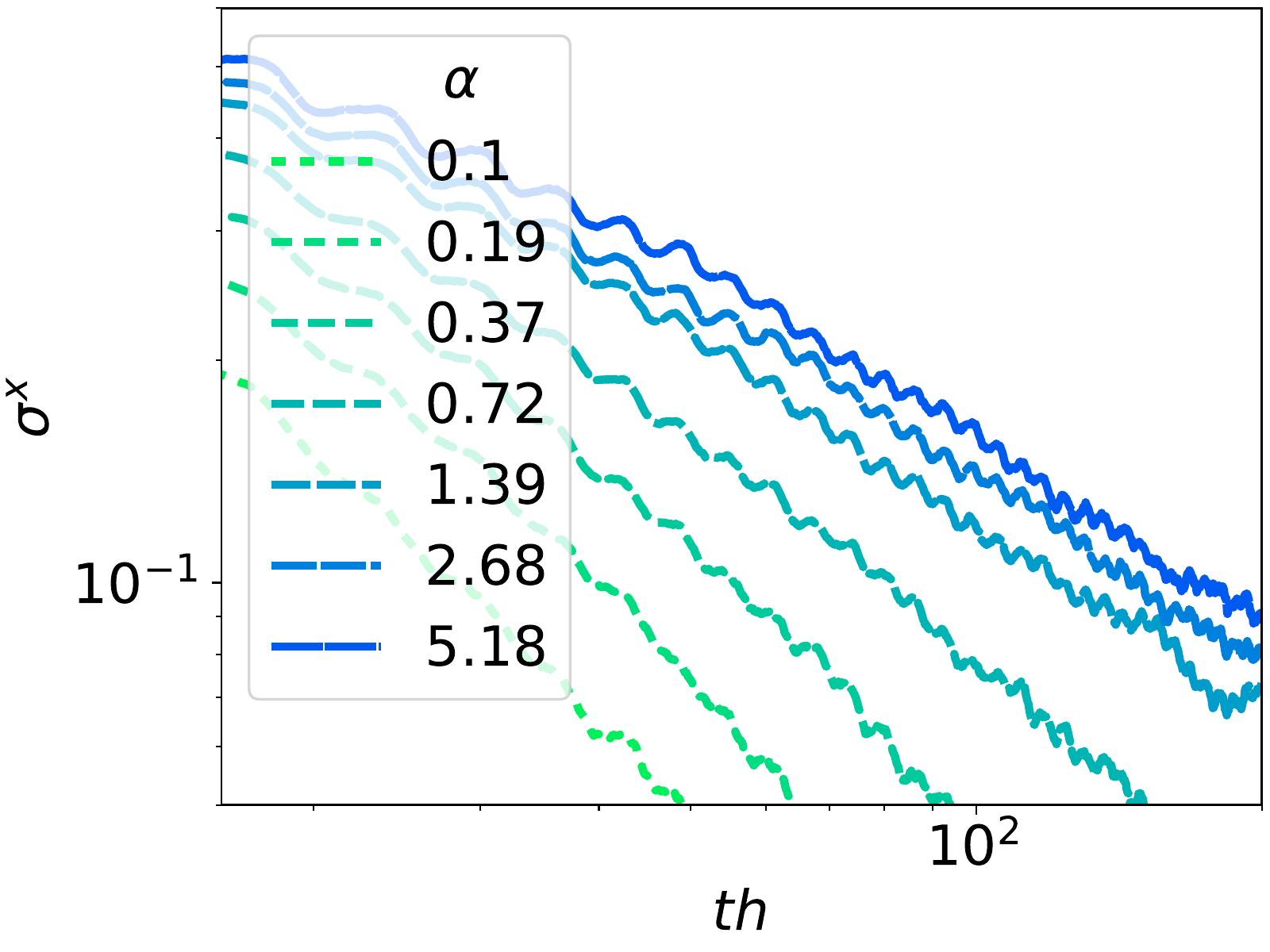}\hspace{20pt}
     \includegraphics[width=0.35\textwidth]{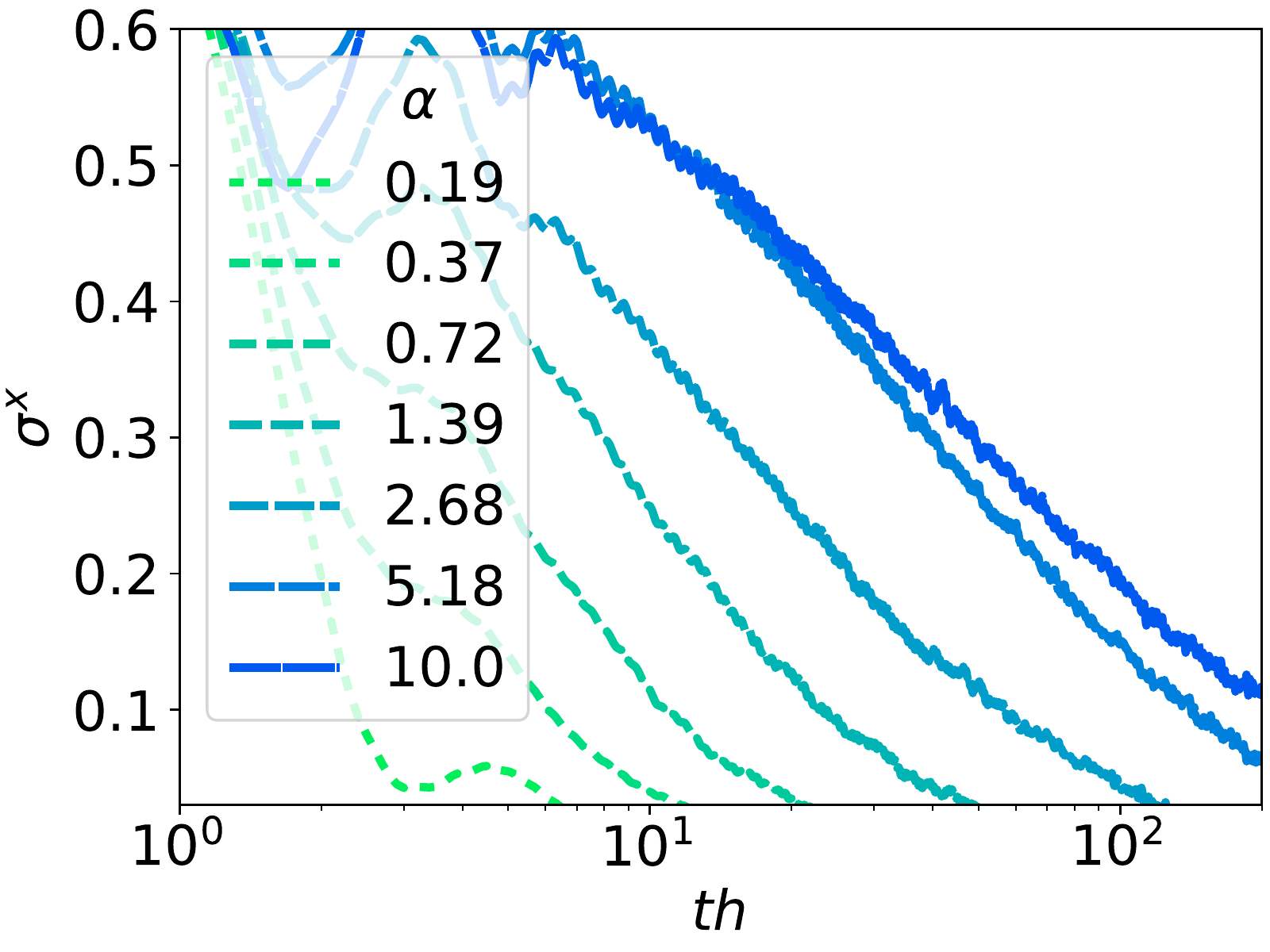}
     \caption{\textbf{Transition from sub-exponential relaxation as a function of $\alpha$.} The \underline{left panel} shows the same data as for the center panel of Fig.~3, but on a ln$(\sigma^x$)-ln$(t h)$ scale. It depicts power-law relaxation for $\alpha>0.72$. The \underline{right panel} shows the same data as the right panel of Fig.~3  on   a $\ln(t h)$ scale. It shows logarithmic-like relaxation for $\alpha>5$.}
     \label{apxfig0}
 \end{figure}

 \begin{figure}[h!]
     \includegraphics[width=0.35\textwidth]{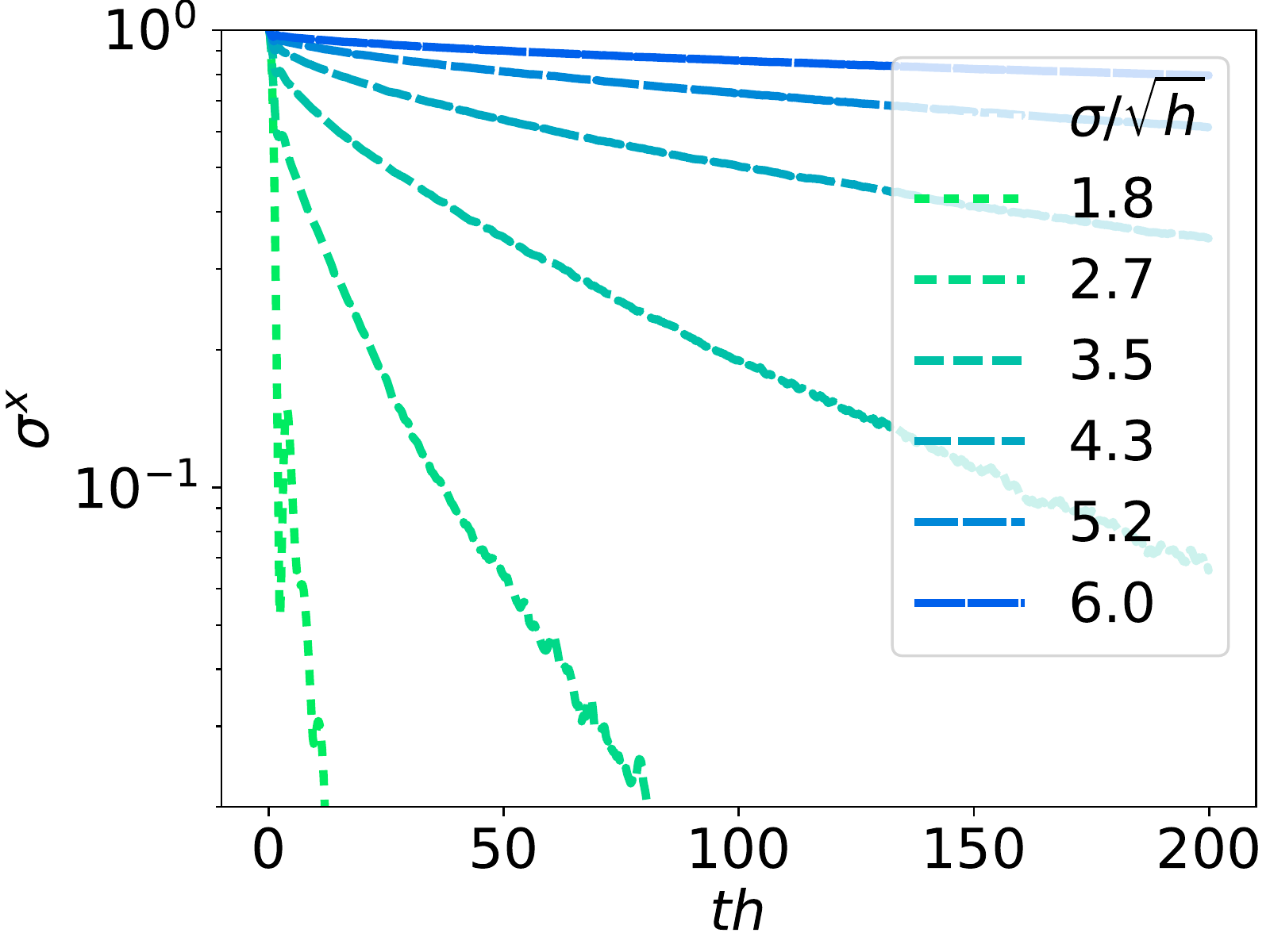}\hspace{20pt}
     \includegraphics[width=0.35\textwidth]{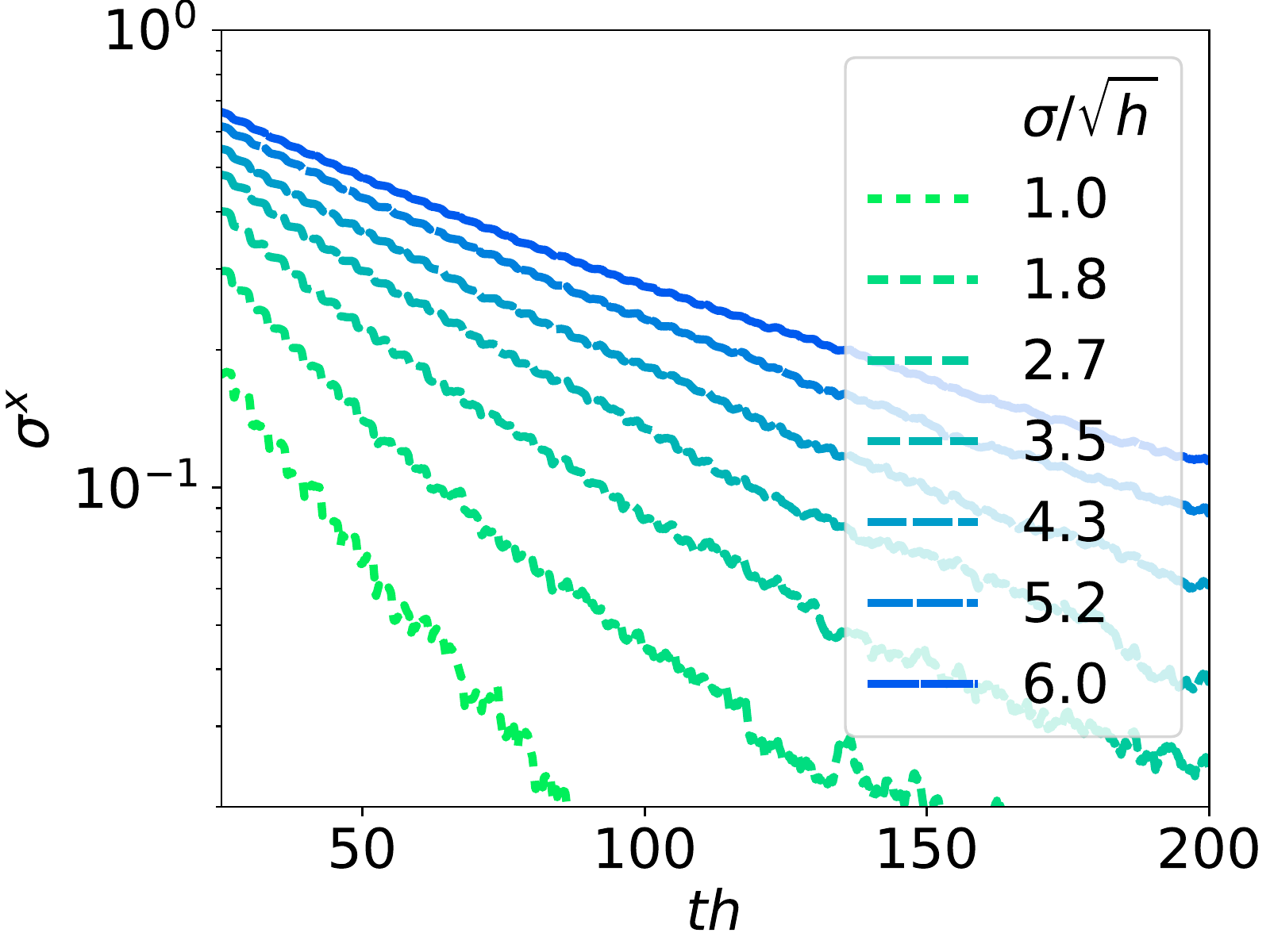}
     \caption{\textbf{Effect of the initial state energy on slow relaxation.} Exponential relaxation   when the energy of the initial state is increased by a shift  the photon amplitude, $\left<a(0)\right>= {2N_s}/{\sqrt{ N_t \omega/h} }$. Both plots are computed for  $\alpha=1$, and $N_t=640$.  The   \underline{left plot} is for $\omega/h=10$;  for the $\sigma/\sqrt{h}\approx2$ line, the $\text{ln}(ht)$ relaxation (Fig.~2 of the main text) turns now into an exponential.  The plot also shows an example (the $\sigma/\sqrt{h}=6$ line) of the region of Fig.~1 labeled 'Slow spins' for large $\omega/h$ and $\sigma/\sqrt{h}$. Here the dynamics are too slow to exhibit a distinguishable relaxation behavior.  The \underline{right plot}  shows similar exponential   relaxation  away from the regime of adiabatic elimination of the photons ( $\omega/h=1$).}
     \label{apxfig0}
 \end{figure}
 \section{Photon Losses}
 To study the effects of photon loss, we add to the photon dynamics Langevin damping and noise~\cite{gardiner2004quantum,gelhausen2018,sels2020a,torre2013}:
 \begin{eqnarray}
     \partial_t a_\lambda =i\omega a_\lambda-\kappa a_\lambda + \kappa \xi_\lambda+\sum_j g_{j\lambda}\sigma^x_{\lambda},
 \end{eqnarray}
 where $\kappa$ is the single photon loss rate and $\xi_i$ is a unit variance Gaussian white noise process.
 In Fig.~\ref{apxfig}, we show that for small $\kappa/h$, the $\ln(t)$ relaxation is maintained for large $\omega$.
 In the same figure, we also show the effect of $\kappa$ in the limit of small $\omega$. Due to the slow convergence of the Euler-Maruyama algorithm~\cite{kloeden1992,kloeden1994,sarkka2019}, we can not access time scales where a power law fit is distinguishable from an exponential or logarithmic fit. We plan to improve numerical convergence at strong loss in forthcoming work.
 \begin{figure}[h!]
     \includegraphics[width=0.35\textwidth]{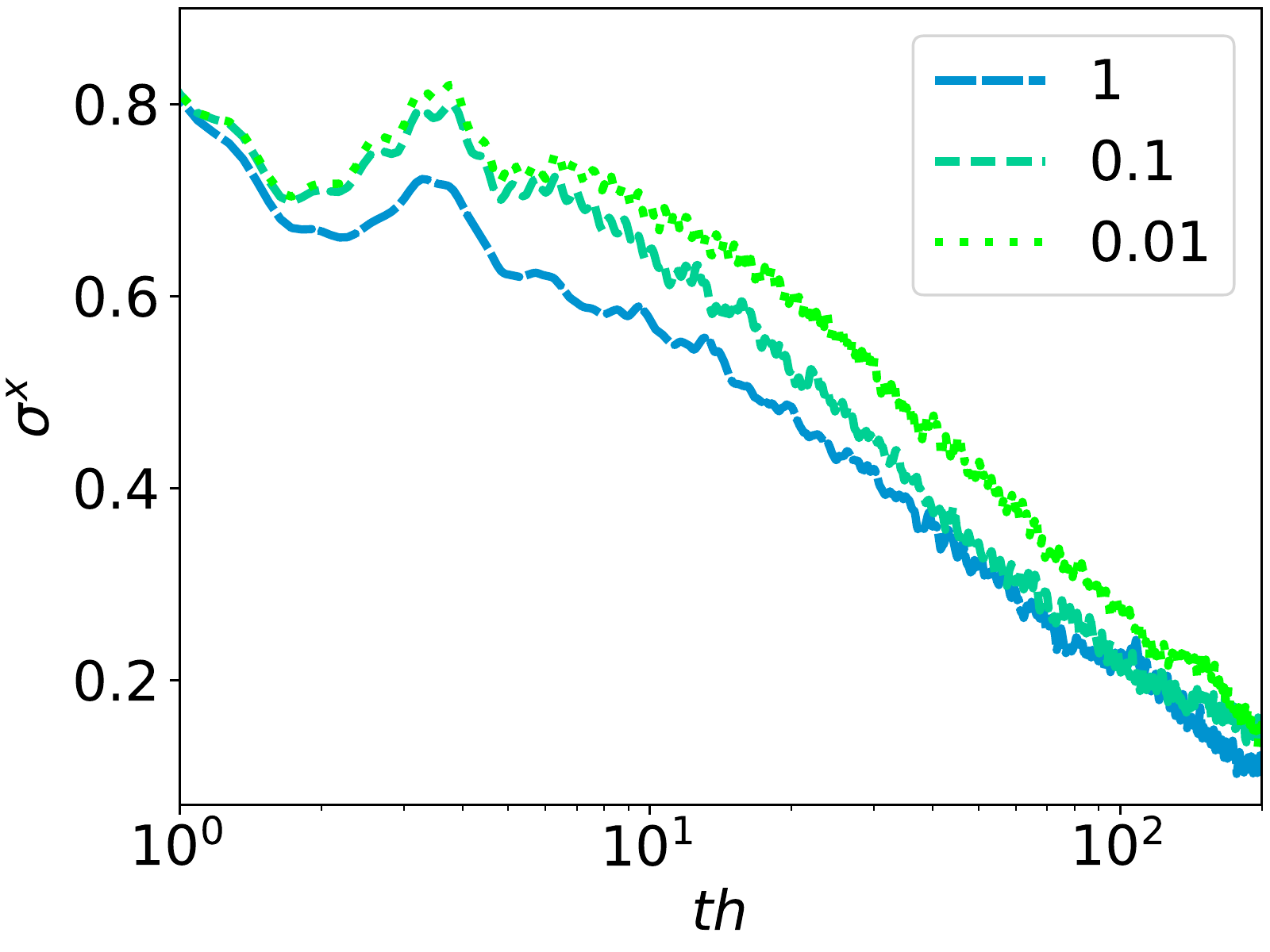}\hspace{20pt}
     \includegraphics[width=0.35\textwidth]{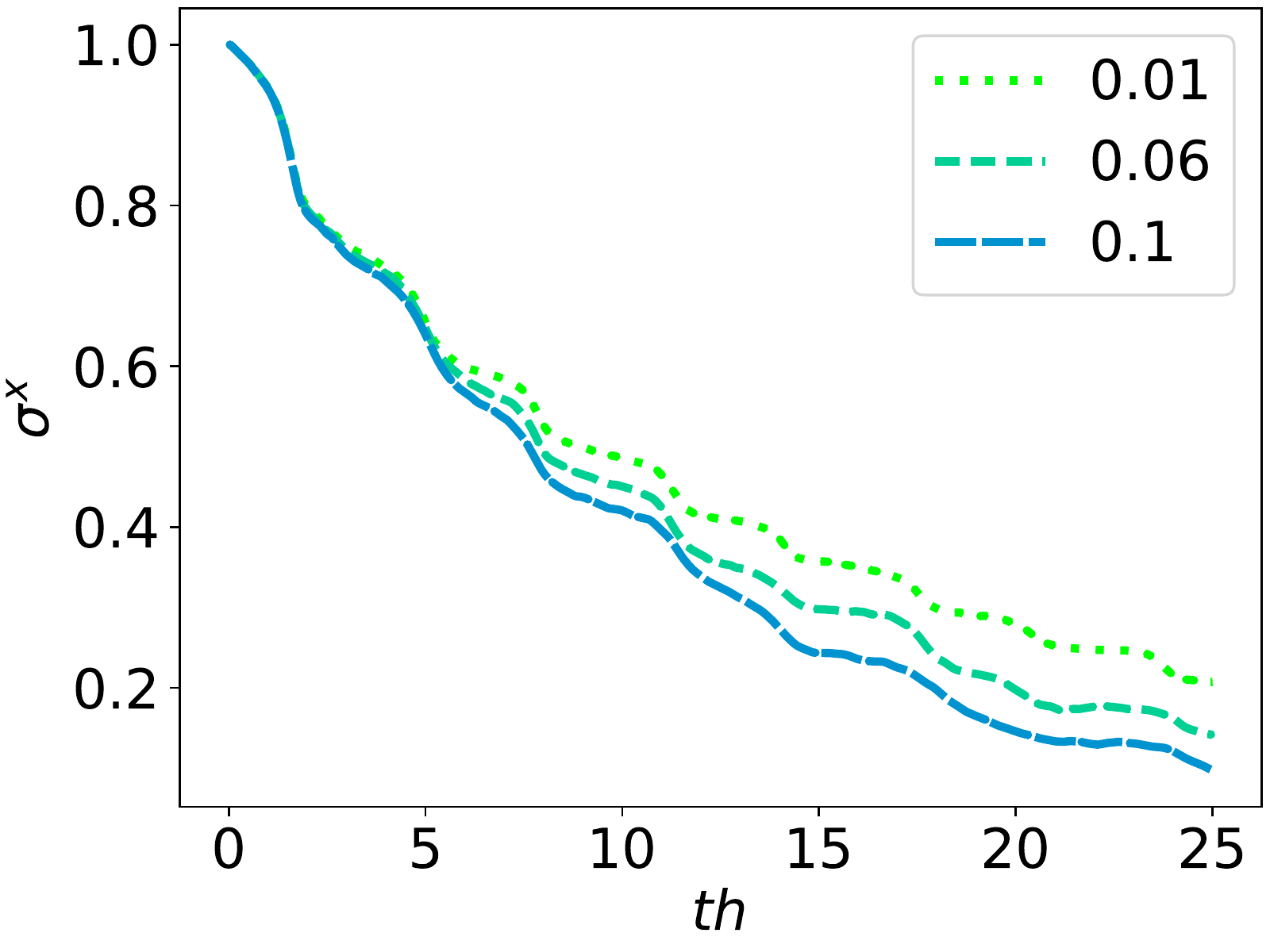}
     \caption{\textbf{Sub-exponential relaxation with dissipation.}  The \underline{left plot} is computed for $\sigma/\sqrt{h}=2$, $\omega/h=10$, $N_t=150$, $\alpha=10$ and with $\kappa/h$ shown in the legend. This plot shows that logarithmic relaxation holds for finite $\kappa$. The \underline{right plot} shows faster relaxation upon increasing losses, in the regime where the adiabatic elimination of the photons breaks down; system's parameters $\sigma/\sqrt{h}=3$, $\omega/h=1$, $\alpha=1$, $N_t=150$ , with $\kappa/h$ shown in the legend.  }
     \label{apxfig}
 \end{figure}

\end{document}